\documentclass[aps,prl,superscriptaddress,showpacs,floatfix,twocolumn,showkeys]{revtex4}

\usepackage{graphicx,epsfig}
\usepackage{dcolumn}
\usepackage{bm}
\usepackage{units}

\begin{document}

\title{Limiting fragmentation of chemical potentials in heavy ion collisions}

\author{
  Laura~A.~Stiles and Michael Murray, University of Kansas, Lawerence, KS 66045-7582\\
 }

\begin{abstract}
Thermal models have been used to successfully describe the  hadron yields from heavy ion 
collisions at a variety of energies. For $\sqrt{s_{NN}} \le 17$ GeV this has usually been done using yields integrated over $4\pi$ but at the higher energies available at RHIC, 
yields measured at central rapidity have been used. 
Recent BRAHMS data allows us to test whether thermal 
models can be generalized to describe the rapidity dependence of particle ratios. We have used the THERMUS package to fit BRAHMS data for the 5\% most
central Au+Au collisions for several rapidities at $\sqrt{s_{NN}} = 62$ and 200 GeV. 
We have found a relationship between the strange and light quark chemical potentials, 
 $\mu_S = 0.21 \pm 0.01 \mu_B$. Using this relation  we are able to 
describe the energy dependence of $\Lambda$, $\Xi$ and 
$\Omega$ ratios from other experiments. We also find that the chemical potentials are consistent with limiting fragmentation. 
\end{abstract}

\maketitle

\section{Introduction}

A wide variety of data from the four RHIC experiments is consistent with the idea that high energy heavy ion collisions produce an almost perfect partonic fluid \cite{WhitePapers}. The hot dense fluid expands in both the transverse and longitudinal directions until the quarks freeze-out into hadrons. In contrast to initial expectations 
these collisions produce particle distributions that  are not boost invariant along the beam axis but rather are Gaussian with widths that depend upon particle type, \cite{BrMeson}. This suggests that the ``chemistry" of the quarks may depend upon rapidity. 
Traditionally thermal analyses have assumed that there is only one source in heavy ion collisions and have used particle yields integrated over all phase space as input to grand canonical fits.
At RHIC several groups have shown that such models can give an excellent description of a large number of particle ratios at  mid-rapidity  
\cite{Becattini:2000jw,Braun-Munzinger:2001ip}. 
The wide rapidity range of BRAHMS allows us to test if there is more than one thermal source. The THERMUS program is an implementation of the Grand Canonical Ensemble in the interactive C++ framework of ROOT 
\cite{Thermus,Root}. THERMUS assumes that at freeze-out all hadrons are in chemical equilibrium.   
Within this model the conservation of isospin, strangeness and baryon number is controlled by chemical potentials and a 
freeze-out temperature T. The model assumes that the total net strangeness is zero and that the isospin of the system is set by the neutron/proton ratio of the colliding beams. While these two things are obviously true for $4\pi$ yields, 
they are assumptions in our analysis.

BRAHMS has measured $\pi^\pm,  K^\pm, p$ and ${\bar p}$  yields versus rapidity at $\sqrt{s_{NN}}=$ ~200 and 62 GeV, \cite{BrStop,BrMeson,BrQM05}
(The 62 GeV data are still preliminary).
From these data it is possible to construct five independent ratios,
$\pi^-/\pi^+, K^-/K^+, K^+/\pi^+, 
{\bar p}/p$ and ${\bar p}/\pi^-$ 
that are input to THERMUS. This is done for data at y=0,1,2 and 3. 
BRAHMS is a spectrometer experiment and so data for all particles of the 
same sign of electric charge is taken at the 
the same magnetic field setting. Therefore some interpolation is required to make ratios at the same rapidity.  

\section{The Grand Canonical Ensemble} 
In a chemical analysis within the grand-canonical ensemble, the statistical-thermal model requires at least five parameters as
input: volume V, the chemical freeze-out temperature T, baryon-, strangeness- and charge chemical potentials $\mu_B, \mu_S$ and $\mu_Q$
respectively. In addition a strangeness saturation factor 
$\gamma_S$ is sometimes employed to allow for the possibility that the system will not have time to equilibrate strangeness. 
When considering particle ratios the volume drops out. 
Since we only have five independent ratios we have to fix some of the parameters ``by hand." For this analysis we fix take T=160 MeV and 
$\gamma_S$ = 1. In addition $\mu_Q$ is constrained to reproduce the proton/neutron ratio for Au+Au collisions. This leaves five ratios to constrain two free parameters, $\mu_S$ and $\mu_B$. 

\begin{figure*}[h]
\epsfig{file=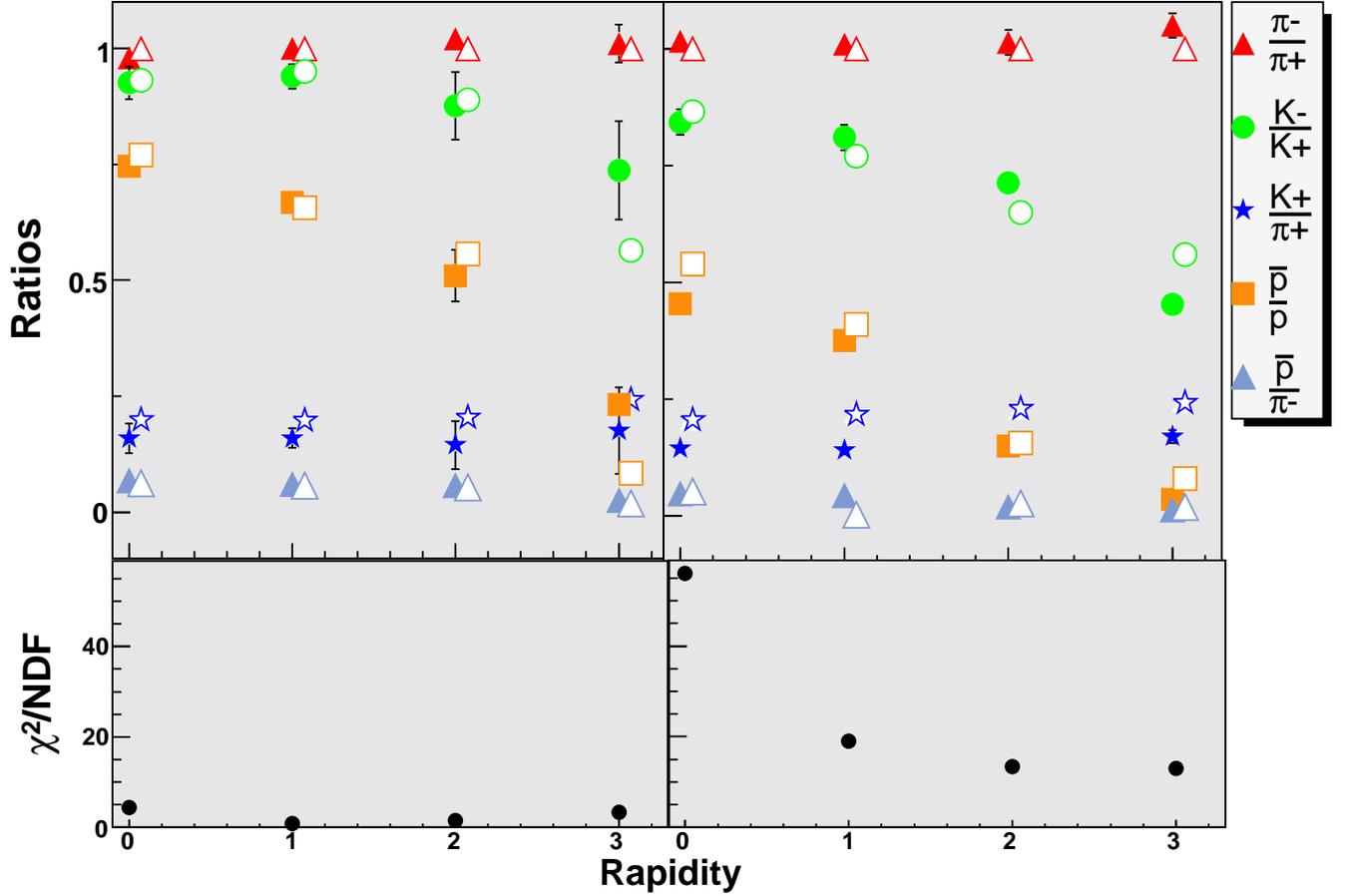,width=\textwidth}
\caption{\label{FitRatios} (Top) The rapidity dependence of our $\pi^-/\pi^+$, $K^-/K^+, K^+/\pi^+, 
{\bar p}/p$ and ${\bar p}/\pi^-$ at $\sqrt{s_{NN}}=$ 200 GeV (left) and preliminary 62 GeV (right). The data are shown by solid symbols and the Thermus fits by open symbols. (Bottom) the $\chi^2$ per degree of freedom. }
\end{figure*}

\section{Fits to BRAHMS data}
Figure \ref{FitRatios} shows fits of particle ratios measured in BRAHMS at 
$\sqrt{s_{NN}}=$62 and 200GeV to THERMUS. The ratios are reasonably close 
but the values of $\chi^2/$NDF are quite large, particularly at $\sqrt{s_{NN}}=$62 GeV. 
This may indicate a failure of the model, an underestimation of the (preliminary) errors or the fact that we only have five particles ratios to constrain the fit. 
For y=0 at $\sqrt{s_{NN}}=$62 GeV, 
the  fit is constrained by the requirement $\mu_S \ge 0$. 
For this point we have
redistributed the probability at negative $\mu_S$ to positive values  using a Gaussian distribution. 
This increases the error on $\mu_S$ for this point. 

\section{Correlations \& Limiting Fragmentation}
\begin{figure}
\epsfig{file=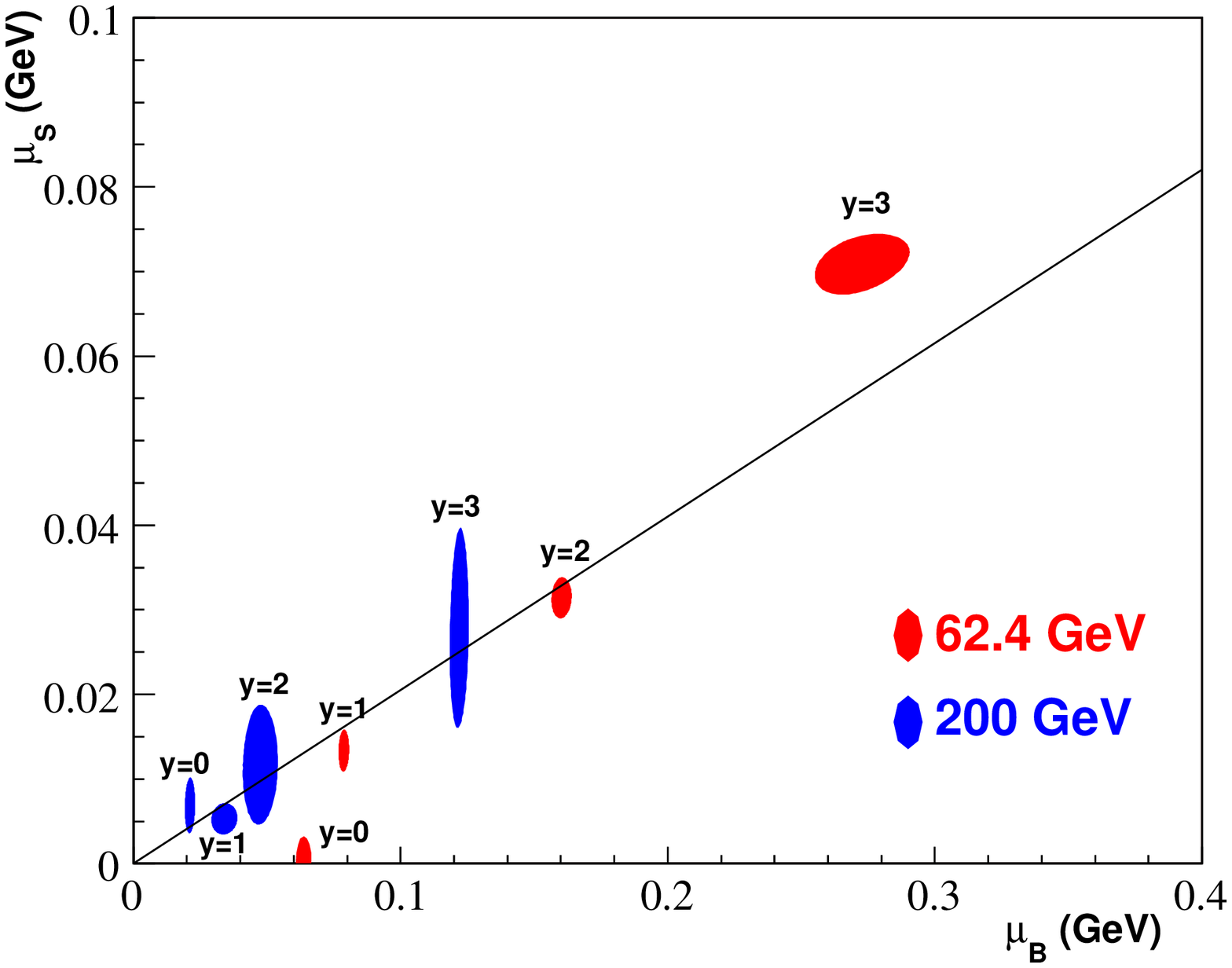,width=\columnwidth}
\caption{\label{MuSMuB} 
The chemical potential of strange quarks $\mu_S$ versus that of light quarks $\mu_B$ for chemical fits to BRAHMS data using THERMUS. The line shows a least squares fit to these results.}
\end{figure}

\begin{figure}
\epsfig{file=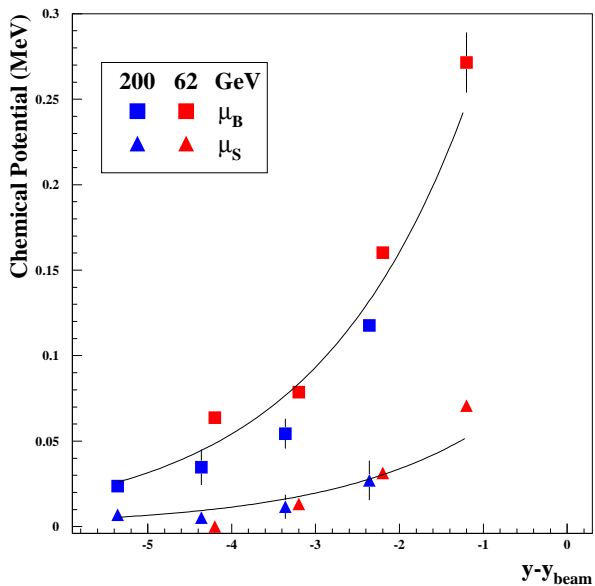,width=\columnwidth}
\caption{\label{muqfrag} 
The chemical potentials versus $y-y_{beam}$. The upper line shows an exponential
fit to the $\mu_B$ data. The bottom line is the same fit scaled by 0.21.}
\end{figure}

 Figure \ref{MuSMuB} shows the correlation between $\mu_S$ and $\mu_B$. 
It is clear that there is a linear relationship between the two potentials.  
We have done a least squares fit and find 
$
\mu_S = (0.21 \pm 0.01) \cdot \mu_B.
$
with a $\chi^2$/NDF = 34.3/7. Most of the $\chi^2$ is generated by the 
y=0 point at 62GeV. 
Figure \ref{muqfrag} plots the results of these fits as a function of $y-y_{beam}$. 
Both chemical potentials rise as a function of $y-y_{beam}$ and at a fixed rapidity 
decrease with energy.  
The upper line shows an exponential fit to the $\mu_B$ data. The data are reasonably consistent with the hypothesis of limiting fragmentation though the 62GeV data are somewhat above the curve while the 200GeV data are below it. The lower line is just 21\% of the upper one. 

\section{Energy dependence of hyperon ratios}
\begin{figure}
  \epsfig{file=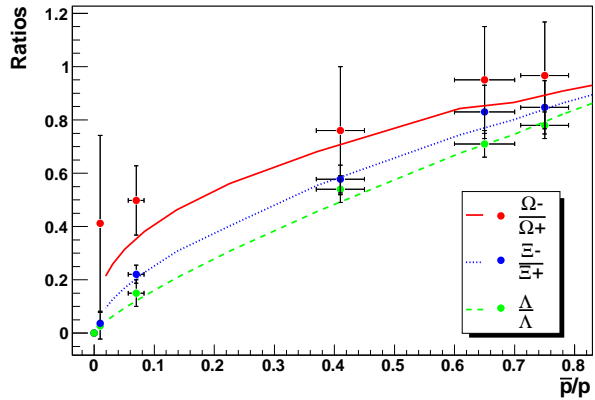,width=\columnwidth}
  \caption{${\bar \Omega}/\Omega$ and 
${\bar \Xi}/\Xi$ ratios from STAR, NA57 and NA49 versus predictions from THERMUS 
using  $\mu_S = 0.21 \cdot \mu_B$
\protect{\cite{NA49Hyperons,NA49Hyperons,StarHyperons}}. The 9, 62 and 200 GeV data are still preliminary.
}
  \label{XiOmega}
\end{figure}

Using this relationship THERMUS can predict the correlations between anti-hyperon/hyperon ratios and the ${\bar p}/p$ ratio since they are controlled, 
by the two fugacities, $e^{-\mu_S/T}$ and  $e^{-\mu_B/T}$ . 
Note for these ratios that any strangeness saturation factor cancels out.) 
In Fig.~\ref{XiOmega}  the  
${\bar \Omega}/\Omega$, ${\bar \Xi}/\Xi$, ${\bar \Lambda}/\Lambda$ ratios  
measured at y=0 for central PbPb and AuAu collisions at $\sqrt{s_{NN}}=$9~-~200 GeV are plotted against
the corresponding the ${\bar p}/p$ ratios from the same experiments
 \cite{NA49Hyperons,NA57Hyperons,StarHyperons}. 
The curves show the corresponding predictions from THERMUS given
$\mu_S = 0.21 \cdot \mu_B.$
The agreement between the prediction and data is good, although THERMUS is slightly below the  
${\bar \Omega}/\Omega$ data.  
Note that to first order the uncertainty on the temperature T does not affect the shape of these curves.

\section{Conclusions} 
Using a Grand Canonical description of heavy ion collisions encoded in the THERMUS framework we can give a rough description of the 
rapidity dependence of hadron ratios from AuAu collisions at $\sqrt{s_{NN}}=$62 and 200 GeV. Given that only
five independent ratios are available there is a limit to how well we can test the thermal ansatz. 
Away from mid-rapidity we cannot prove that T=160MeV or that there is zero net strangeness in each unit of rapidity. 
However 
if one accepts this ansatz several trends are evident. 
The chemical potentials of both light and strange quarks increases with rapidity and decrease with energy, 
implying that different regions of rapidity lose contact with each other before reaching chemical equilibrium. 
The chemical potentials seem to obey limiting fragmentation and rise exponentially with $y-y_{beam}$. 
Finally the strange quark chemical potential is proportional to the light quark chemical potential, 
$\mu_S = (0.21 \pm 0.01) \cdot \mu_B.$ Using this relation 
 we are able to describe the energy dependence of hyperon production.

\begin{acknowledgments}
We wish to thank Spencer Weaton for help with the THERMUS package and Sevil Salur for many useful discussions. 
Work supported  
by the DOE Office of Science under
contracts DE-FG03-96ER40981, EPSCoR
 DE-FG02-04ER46113 and the Kansas Technology Enterprise Corporation. 
\end{acknowledgments}

\bibliographystyle{apsrev}
\bibliography{apsrev}

\end{document}